\documentclass[prl,twocolumn,floatfix]{revtex4}
\usepackage[dvips]{graphicx}

\begin{document}

\title{Corner transfer matrix renormalisation group method for 
two-dimensional self-avoiding walks  and other $O(n)$ models}

\author{D P Foster}
 \author{C Pinettes}
\affiliation{Laboratoire de Physique Th\'eorique et Mod\'elisation
(CNRS UMR 8089), Universit\'e de Cergy-Pontoise, 5 mail Gay Lussac
95035 Cergy-Pontoise cedex, France}

\begin{abstract} 
We present an extension of the corner transfer matrix renormalisation
group (CTMRG) method to $O(n)$ invariant models, with particular interest in
the self-avoiding walk class of models ($O(n=0)$). The method is
illustrated using an interacting self-avoiding walk model. 
Based on the
efficiency and versatility  when compared to other
available numerical methods, we present 
CTMRG as the method of choice for 
two-dimensional self-avoiding walk problems.
\end{abstract}

\maketitle


The self-avoiding walk class of models on the two-dimensional square
lattice, with a variety of possible interactions, has mobilised the
scientific community for about half a century\cite{clois,vander}. 
The number of exact
results for such models is limited, and numerical studies
are hard. A clear illustration of the numerical difficulty 
is the disagreement
which existed
over the numerical determination of critical temperature and exponents
for the standard $\theta$-point model, see for example
references\cite{thetanum}.

To date the numerical methods available for the study of interacting
self-avoiding walks in two dimensions are series expansions of walks of
lengths of a few tens of steps\cite{enumer}, 
transfer matrices for lattice widths
up to about 12\cite{derr} 
and increasingly complicated Monte-Carlo simulation
methods\cite{mcslow}, limited in practice to only a portion of the phase
diagram. 

Motivated by these numerical difficulties, 
we decided to extend the Corner
Transfer Matrix Renormalisation Group (CTMRG) method\cite{nishino}. 
The CTMRG method is based on  White's Density Matrix
Renormalisation Group method (DMRG)\cite{white} and Baxter's corner
matrix formalism\cite{baxter}. 
To date the
CTMRG method has only been applied to discrete spin models, where it is
shown to be computationally efficient\cite{nishino}. 

Our
extension to interacting self-avoiding walk models exploits the
connection betweeen these models and the $O(n)$ invariant spin
models\cite{nienon}, which contains as special cases the
Ising model ($n=1$), the XY model ($n=2$) and the Heisenberg model ($n=3$).
The method therefore has applications well beyond the
self-avoiding walk type models ($n=0$). 

The $O(n)$ spin model is defined through the partition function\cite{on}
\begin{equation}\label{onpart}
{\cal Z}_{O(n)}=\sum_{\{\vec{s}_i\}} \exp\left(\frac{1}{2}\beta
J\sum_{\langle i,j \rangle} \vec{s}_i\cdot\vec{s}_j\right),
\end{equation}
where $\langle i,j\rangle$ refers to a sum over nearest
neighbour spins. The spin $\vec{s}_i$ has $n$ components, and is
normed such that $s^2_i=1$. 
Another formulation of $O(n)$ invariant models, with the same critical
behaviour, is:
\begin{equation}\label{nonpart}
{\cal Z}_{O(n)}=\sum_{\{\vec{s}_i\}} \prod_{\langle i,j\rangle}
\left(1+K\vec{s}_i \cdot\vec{s}_j\right).
\end{equation}
where the spins are now placed on the lattice bonds\cite{nienon}.
A diagrammatic
 expansion of Equation~\ref{nonpart}
follows if we identify the $1$ as the weight of an empty bond between 
the sites $i$ and $j$ and the $K$ as the weight of an occupied bond. 
This expansion may be expressed in terms of graphs $\cal G$ of 
non-intersecting loops (collisions at sites
are however allowed)\cite{nienon}.
The partition function may then be written:
\begin{equation}\label{gpart}
{\cal Z}_{O(n)}=\sum_{{\cal G}} n^{l({\cal G})} K^{b({\cal G})},
\end{equation}
where $l$ is the number of  loops and $b$ is the number
of occupied bonds. 

The parameter  $n$ is now a fugacity controlling the number of
loops, and  need no longer be taken as an
integer. This fugacity corresponds to a long ranged interaction,
since the loops may be of any size. This non-locality is
undesirable for our purposes. We would like to express $n$ as a
product over local weights.
This may be achieved as follows. 
Each loop may
be followed clockwise or anticlockwise.
The loops in
Equation~\ref{gpart} are
 not oriented, but may be oriented
by associating $2^{l({\cal G})}$ oriented graphs with each
non-oriented graph.
By associating a
loop fugacity $n_+$ ($n_-$) with the clockwise (anticlockwise) 
oriented loops, 
the partition function may
be rewritten\cite{nien2}:
\begin{equation}
{\cal Z}_{O(n)}=\sum_{\cal G} (n_+ + n_-)^l K^b =\sum_{{\cal
G}^{\prime}}n_+^{l_+} n_-^{l_-} K^b 
\end{equation}
where ${\cal G}^\prime$ is the set of oriented loop graphs, and $l_+$ ($l_-$)
is the number of clockwise (anticlockwise) oriented loops. 
Setting $n_+=\exp(i\theta)$ and
$n_-=\exp(-i\theta)$ gives $n=2\cos(\theta)$.
The
oriented loop factor is now broken up into local weights 
by associating a corner weight $w_i=\exp(i\theta/4)$ with every
clockwise corner and $w_i=\exp(-i\theta/4)$ with every anticlockwise
corner.
On the square lattice, there must
be four more corners with one orientation, compared with the other
orientation, in order to close a loop. The product of the local
weights will then give the correct weight for the oriented loops.

\begin{figure}
\begin{center}
\includegraphics[width=7cm]{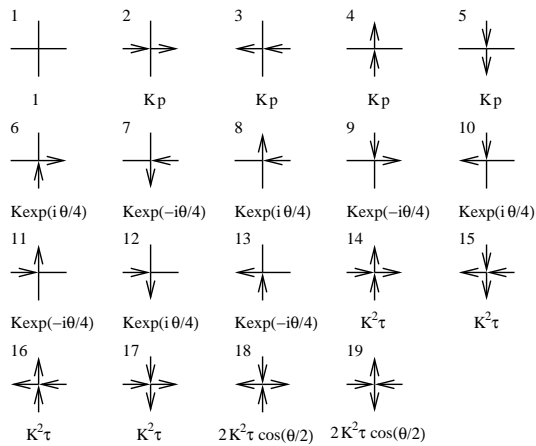}
\end{center}
\caption{The 19 allowed vertices in the most general $O(n)$ model.
$K$ is the step fugacity and $\tau=\exp(-\beta\epsilon)$, where
$\epsilon$ is the attractive monomer-monomer interaction energy.}
\label{verts}
\end{figure}

The partition function may now be rewritten in terms of a vertex
model\cite{nienon}:
\begin{equation}\label{vertpart}
{\cal Z}_{O(n)}=\sum_{{\cal G}^\prime} \prod_i v_i
\end{equation}
where $v_i$ is the weight of the vertex at site $i$.  
The derivation
given here is only for the simplest case, but we may freely change
the weights of the vertex configurations in order to generate
different interactions in the original model (see figure~\ref{verts}).

The limit $n\to 0$ corresponds to the self-avoiding walk model\cite{pgdg}.  
In the generalised form presented here this
corresponds to an interacting self-avoiding walk model
due to Bl\"ote and Nienhuis\cite{nienon,blbatnien}. A
 step fugacity $K$ and an
attractive short ranged interaction $\varepsilon<0$ are introduced. 
In the
standard $\theta$ point model the interactions are between
non-consecutively visited nearest-neighbour sites and a given site may
only be visited once\cite{clois,vander}. In our current models, this last constraint is
relaxed; the walk may collide at a site, but not cross, and remains
self-avoiding for the bonds. The interaction is now assigned to doubly
visited sites. An additional weight $p$ is added for sites which are
visited by a straight section of walk (i.e. do not sit on a corner). 
The partition function may then be written:
\begin{equation}
Z=\sum_{\rm walks} (Kp)^L\tau^{N_I}p^{-N_c},
\end{equation}
where $\tau=\exp(-\beta\varepsilon)$, $N_I$ is the number of site
collisions and $N_c$ are the number of corners in the walk. This model
gives rise to the vertex weights shown in figure~\ref{verts}.
The standard self-avoiding walk model is found setting $\tau=0$ and
$p=1$ and when $p=0$ the model has
the same critical behaviour as the standard $\theta$ point
model\cite{blbatnien}. 
During the remainder of this letter we shall illustrate the CTMRG
method for $O(n)$ models using this Bl\"ote-Nienhuis walk model.

Following Baxter\cite{baxter},
the partition function of a two dimensional
lattice model may be written in terms of the product of four
matrices representing the four quarters of the lattice.  
The inputs and outputs of the matrices are
the configurations at the seams of the four quarters. These
matrices are known as corner transfer matrices. In general the
four matrices are different, but may often be related by lattice
symmetries. For our model the four matrices are the same up to a
complex conjugation operation\cite{preprint}.

It is usually not possible to  calculate explicitly these
matrices for systems with a large number of sites. This is where
the CTMRG method comes in;
the matrices for larger lattices are
calculated from smaller lattices iteratively\cite{nishino}.  
This is done as
follows. An initial system, consisting of a small number of sites, is
mapped exactly onto a prototype system made up of four $m$-state spins.
At
each iteration the system is enlarged by adding sites, this
enlarged system is then projected back onto the prototype system
in some optimal way, so as to minimise the loss of information.
The value of $m$
determines the amount of information which may be carried forward
at each iteration, the larger the value of $m$ the better the
approximation. For details see \cite{nishino,preprint}.


\begin{figure}
\begin{center}
\textbf{(A)}

\includegraphics[width=8cm]{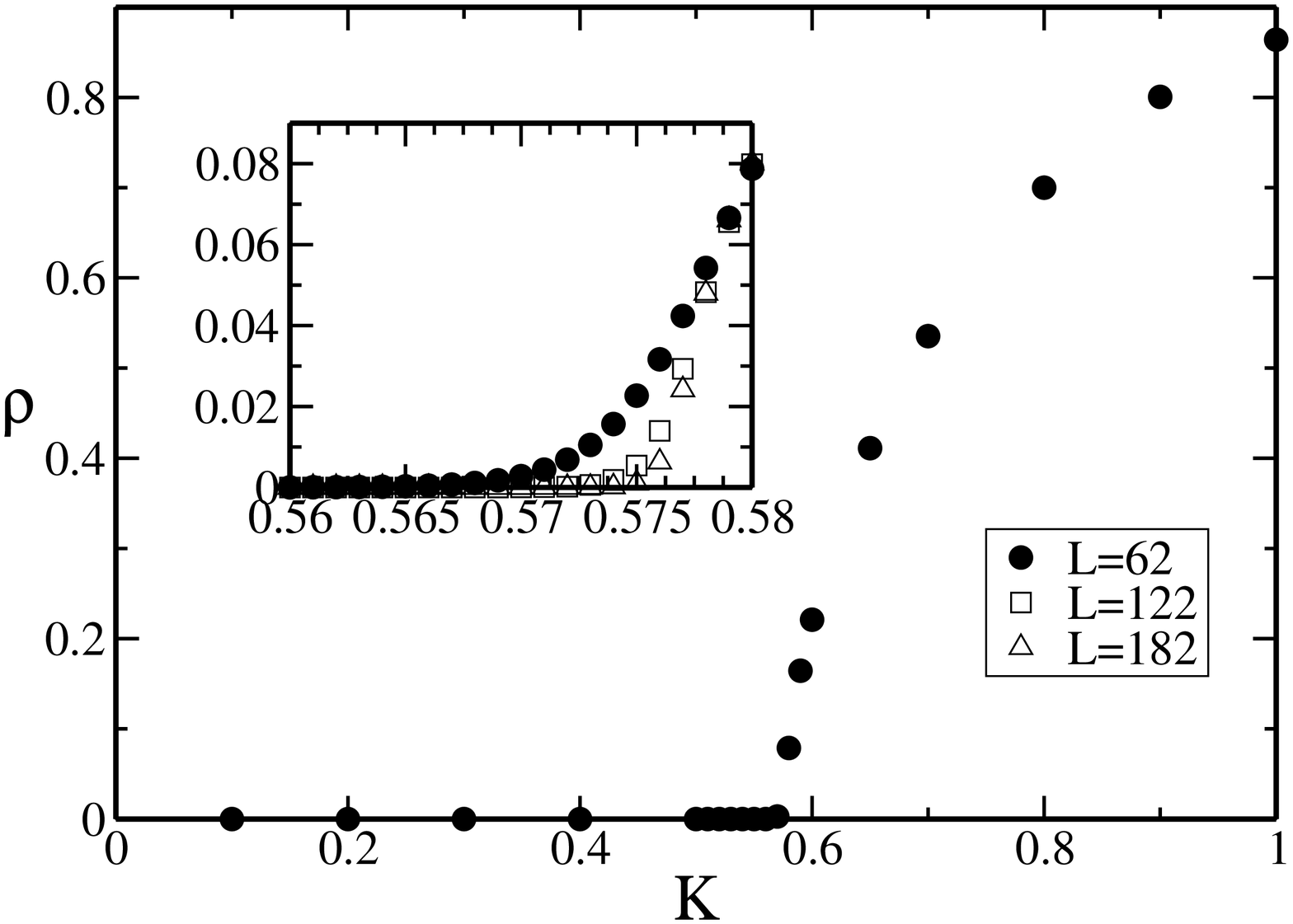}

\textbf{(B)}

\includegraphics[width=8cm]{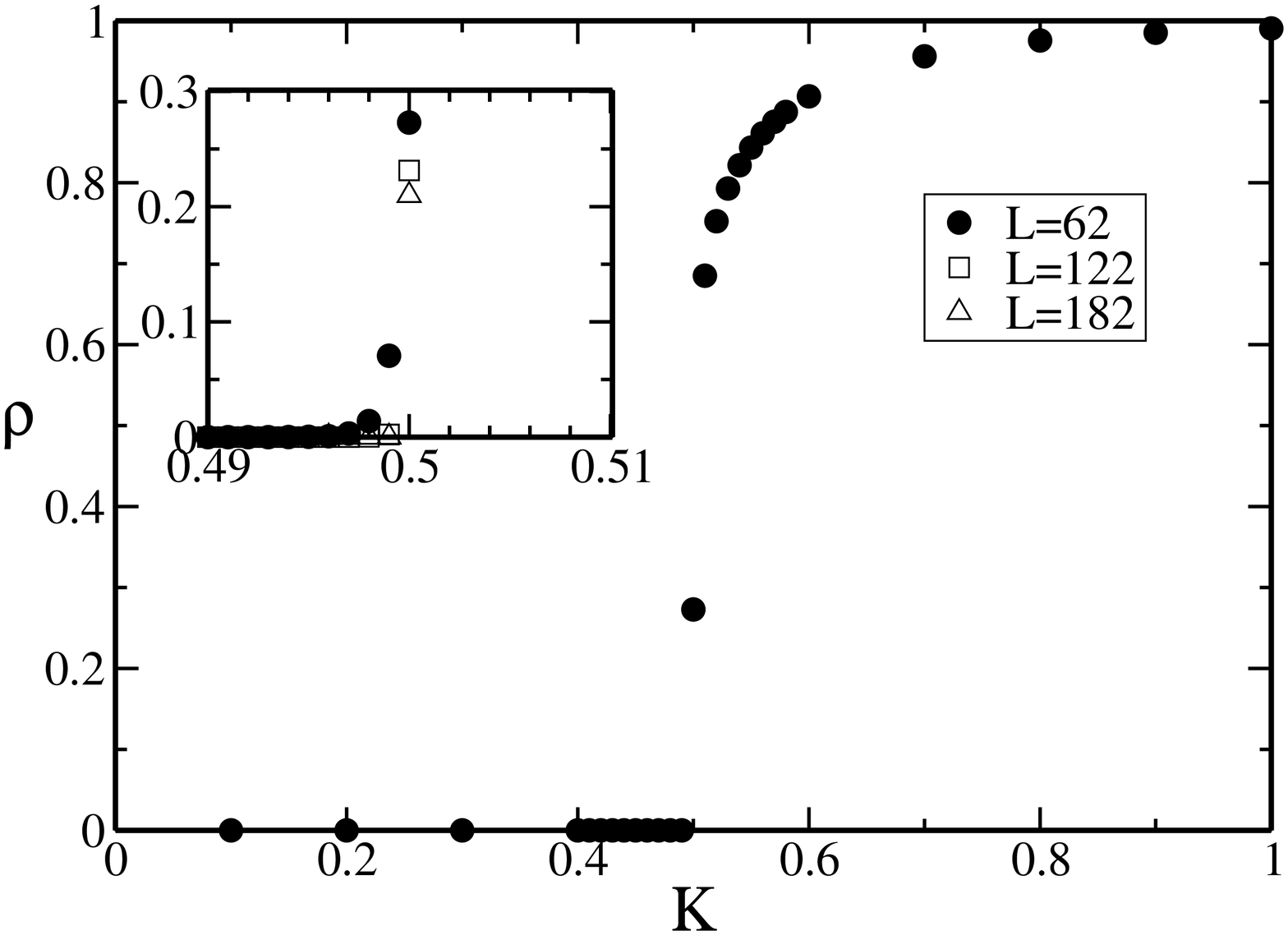}

\textbf{(C)}

\includegraphics[width=8cm]{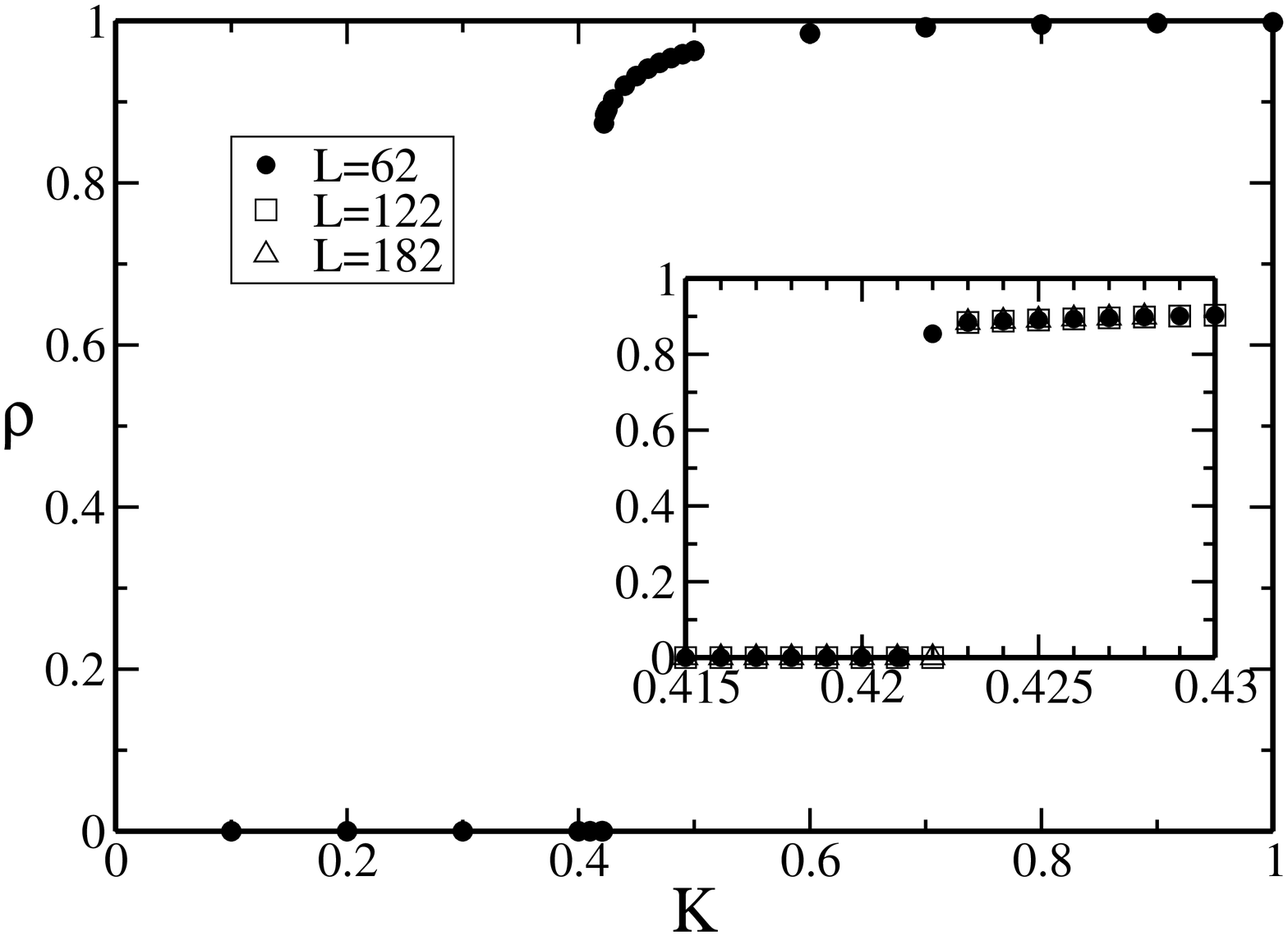}
\end{center}
\caption{The density, $\rho$, as a function of $K$ with $p=0$ and
(A) $\tau=1$, (B) $\tau=2$ and (C) $\tau=3$ for $L=62, 122$ and 182. 
Estimates of $K_c$ from finite size scaling are given in
table~\ref{table}. The value of $K$ for the first order transition when
$\tau=3$ estimated directly from the jump in the density was
found to be $K^*=0.422\pm 0.001$ (compared to $0.421\pm 0.001$ found
previously\cite{blbatnien}).
}\label{plot} 
\end{figure}

As with the DMRG method, the innermost sites of the lattice are
treated exactly. This means that the CTMRG method is most appropriate
for the calculations of one-point functions such as the site free
energy, the site density of monomers, specific heats etc. In
figure~\ref{plot} 
we show the density, $\rho$, as
a function  of the step-fugacity $K$ for fixed values of $\tau$.
We chose to fix $p=0$ since for this value the Nienhuis-Bl\"ote model
has the same critical behaviour as the standard $\theta$ point
model\cite{blbatnien}.  
When $\tau<2$ the model is expected to be in the
self-avoiding walk universality class and present a critical
transition where the density changes continuously from
$\rho=0$ for $K<K_c$ to $\rho>0$ for $K>K_c$. When
$\tau=2$ the transition is expected to be tricritical, in the same
universality class as the $\theta$ point, with $K_c=1/2$
exactly. For $\tau>2$ the transition is expected to be first
order. These three types of behaviour may be clearly seen in
figure~\ref{plot}.

Whilst the raw finite-size data presented in figure~\ref{plot} already
gives a fairly precise idea of the behaviour of the system (order of
the transition and first estimate of $K_c$), 
practical calculations of critical temperature  and critical exponent
estimates 
rely on finite-size analysis. In transfer
matrix calculations one usually uses
a phenomenological renormalisation group based on the correlation
lengths due to
Nightingale\cite{mpn76}.
In the CTMRG method an 
effective transfer matrix may be determined directly.
However, as already mentioned, 
the highest precision is obtained for
the one point correlation functions. Since
the number of sizes available is an order of magnitude larger than
that available for transfer matricies 
it is advantageous to exploit the finite size
scaling laws for the one point functions.
In particular it is 
expected that the singular part of the density scales as
\begin{equation}\label{scalingfn}
\rho_s(K,L)=L^{1/\nu-2}\tilde{\rho}(|K-K_c|L^{1/\nu}).
\end{equation}
This scaling behaviour implies that the function
\begin{equation}
\varphi_{L,L^\prime}(K) =
\frac{\log\left(\rho_s(K,L)/\rho_s(K,L^\prime)\right)}{\log\left(L/L^\prime
\right)}
\end{equation}
takes the value
$\varphi_{L,L^\prime}=1/\nu-2$ when $K=K_c$,
independantly of $L$ and $L^\prime$. 
Naturally there are additional
finite size corrections which should be taken into account, but the
conclusion is that if the function $\varphi_{L,L^\prime}(K)$ is
plotted for various values of $L$ and $L^\prime$ then it will converge
to a fixed point given by $\varphi(K_c)=1/\nu-2$. In what follows we
have set $L^\prime=L-2$ and looked for solutions of the equation
\begin{equation}
\varphi_{L,L-2}(K^L_c)=\varphi_{L-2,L-4}(K^{L}_c),
\end{equation}
where $K_C^L$ is the estimated critical temperature from sizes $L$,
$L-2$ and $L-4$. If such a solution does exist then
$K_c=\lim_{L\to\infty}K_c^L$ 
and
$\nu=\lim_{L\to\infty}1/(\varphi_{L,L-2}(K^L_c)-2)$.

\begin{figure}
\center{\includegraphics[width=8cm]{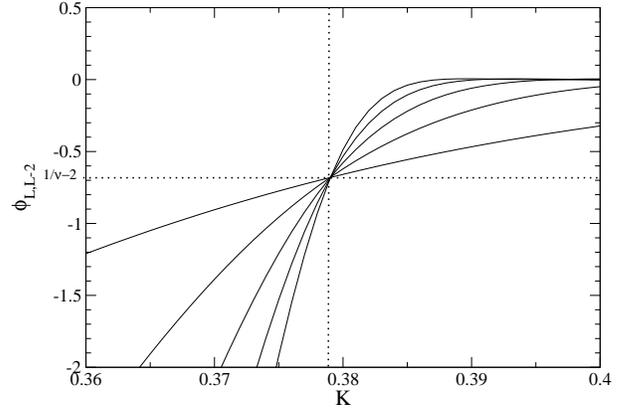}}
\caption{$\varphi_{L,L-2}$ as a function of $K$ for the self-avoiding walk
model ($p=1$, $\tau=0$) for $L=10, 20, 30, 40$ and 50.
The horizontal and vertical lines give the corresponding finite size
estimates of $K_c$ and $\nu$.}\label{sawphi}
\end{figure}

\begin{figure}
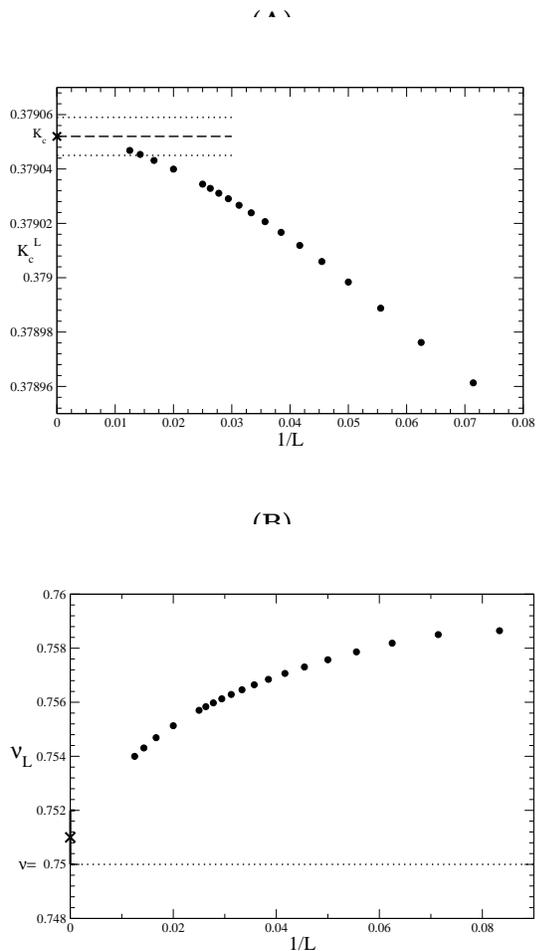


\textbf{(A)}

\ 

\ 

\includegraphics[width=7cm]{fig4a.ps}

\

\

\textbf{(B)}

\

\ 

\includegraphics[width=7cm]{fig4b.ps}
\caption{(A) Critical point and (B)  critical
exponent estimates for the self-avoiding walk ($p=1$, $\tau=0$) as a
function of $1/L$. The extrapolated values of $K_c$ and $\nu$ are
shown. In (A) the three horizontal lines show the previous transfer
matrix estimate for $K_c$ along with the corresponding
error bars\cite{vander}. 
In (B) the horizontal dotted line indicates the exact value of
$\nu=3/4$.} 
\label{sawkc}
\end{figure}

Figure~\ref{sawphi} shows $\varphi$ plotted for several values of $L$
for the pure self-avoiding walk model ($p=1$, $\tau=0$).
The different curves cross at a point defining $K_c$ and $\nu$.
The corresponding estimates of $K_c$ and $\nu$ are shown in 
Figure~\ref{sawkc} plotted as a function of $1/L$. The extrapolations
of $K_c^L$ and $\nu^L$ to $L\to \infty$ are given in Table~\ref{table}
along with prelimenary estimates for $p=0$ and different values of
$\tau$. The accuracy with which the critical points are determined is
in general an order of magnitude better than found with transfer
matricies. Full details and definitive estimates will be given
elsewhere~\cite{preprint}.

\begin{table}
\begin{tabular}{|c|c|c|c|c|}\hline
$p$ & $\tau$ & TM $K_c$  & CTMRG $K_c$ & $\nu$\\\hline
0& 0 & $0.63860\pm 0.00005$& $0.63865\pm 0.00005$& $0.755\pm 0.007$\\
0& 1 & $0.5769\pm 0.0001$ & $0.57686\pm 0.00002$ & $0.74\pm 0.02$\\
0& 2 & $0.5001\pm 0.0001$ & $0.500000\pm 0.000001$ & $0.571\pm 0.001$\\
1& 0 & $0.379052\pm 0.000007$ & $0.379052 \pm   0.000001$ & $0.751\pm0.001$\\\hline 
\end{tabular}
\caption{Estimates of $K_c$ from transfer matrix (TM)
calculations taken from reference \cite{blbatnien} for $p=0$ and
from reference\cite{vander} for the self-avoiding walk ($p=1$ and
$\tau=0$) are compared to our (previsional) estimates using the
CTMRG. First estimates of $\nu$ using CTMRG are also given.}\label{table} 
\end{table}

In (numerically) exact methods, such as the transfer matrix method or
the exact
enumeration method,  the main source of uncertainty in the
results is due to the extrapolation of a small number of
points. In the CTMRG method the extrapolation problems are largely
removed  since lattice sizes an order of magnitude larger have been
reached.
The uncertainty now lies in the precision
related to the calculation of each density point.

In this  article 
results for the Bl\"ote-Nienhuis interacting
self-avoiding walk are presented since good quality results already
exist, providing  a good test of the
efficiency of the CTMRG method. 
In particular we have focused on the self-avoiding walk model ($p=1$, $\tau=0$)
for which the asymptotic limit is well described by
transfer matrix calculations  and
yet we still find a substantial increase in precision (see
Table~\ref{table}).
It is clear that in circumstances where larger system sizes are
required to extract the scaling behaviour, CTMRG should far exceed
the numerically exact methods in performance.

The quality of results presented in this article is virtually unattainable
with such ease 
by any other numerical method  we know of (there is a very small number of
exceptional cases  where better accuracy was
obtained\cite{exceptional}) 
and so we present  CTMRG
as the method of choice for two-dimensional self-avoiding walk
models. CTMRG may easily be extended to more complicated interacting
self-avoiding walk models, such as the hydrogen-bonding self-avoiding
walk\cite{fs}.

\end{document}